# An algebraic method for solving the inverse problem of quantum scattering theory

**N.A. Khokhlov**

Department of higher mathematics, Southwest State University, Kursk, Russia

E-mail: nikolakhokhlov@yandex.ru

**Abstract**

We present a new algebraic method for solving the inverse problem of quantum scattering theory based on the Marchenko theory. We applied a triangular wave set for the Marchenko equation kernel expansion in a separable form. The separable form allows a reduction of the Marchenko equation to a system of linear equations. For the zero orbital angular momentum, a linear expression of the kernel expansion coefficients is obtained in terms of the Fourier series coefficients of a function depending on the momentum $q$ and determined by the scattering data on the finite range of $q$.

## 1. Introduction

The inverse problem of quantum scattering is essential for various physical applications such as the internuclear potential extraction from scattering data and similar problems [1]. The main approaches to solving the problem are Marchenko, Krein, and Gelfand-Levitan theories [2-6]. The ill-posedness of the problem complicates its numerical solution. The development of robust methods for solving the inverse problem remains a fundamental challenge for applications.

This paper considers a new algebraic method for solving the inverse problem of quantum scattering theory derived from the Marchenko theory. To this end, we propose the Marchenko equation's numerical solution in the case of the integral kernel expressed as a separable series in the triangular wave set. Then we show that the series coefficients may be calculated directly from the scattering data on the finite segment of the momentum axis. We offer an exact relationship between the potential function accuracy and the range of known scattering data.

## 2. Theory

We write the radial Schrödinger equation in the form:

$$\left(\frac{d^2}{dr^2} - \frac{l(l+1)}{r^2} - V(r) + q^2\right)\psi(r,q) = 0. \quad (1)$$

Initial data for the Marchenko method [1] are:

$$\{S(q), (0 < q < \infty), \tilde{q}_j, M_j, j = 1, ..., n\}, \quad (2)$$

where $S(q) = e^{2i\delta(q)}$ is a scattering matrix dependent on the momentum $q$. The S-matrix defines asymptotic behavior at $r \to +\infty$ of regular at $r = 0$ solutions of (1) for $q \geq 0$; $\tilde{q}_j^2 = E_j \leq 0$, $E_j$ is $j$-th bound state energies ($-i\tilde{q}_j \geq 0$); $M_j$ is $j$-th bound state asymptotic constant. The Marchenko equation is a Fredholm integral equation of the second kind:

$$F(x,y) + L(x,y) + \int_x^{+\infty} L(x,t)F(t,y)dt = 0. \quad (3)$$

$$F(x, y) = \frac{1}{2\pi} \int_{-\infty}^{+\infty} h_l^+(qx)[1 - S(q)] h_l^+(qy) dq$$

We write the kernel function as $+ \sum_{j=1}^{n_b} h_l^+(\tilde{q}_j x) M_j^2 h_l^+(\tilde{q}_j y)$

$$= \frac{1}{2\pi} \int_{-\infty}^{+\infty} h_l^+(qx) Y(q) h_l^+(qy) dq, \qquad (4)$$

where

$$Y(q) = \left[ 1 - S(q) - i \sum_{j=1}^{n_b} M_j^2 (q - \tilde{q}_j)^{-1} \right] \qquad (5)$$

Solution $L(x, y)$ of eq. (3) gives the potential of eq. (1):

$$V(r) = -2 \frac{dL(r,r)}{dr} \qquad (6)$$

There are many computational approaches for the solution of Fredholm integral equations of the second kind. Many of the methods use an equation kernel's series expansion [7-19]. We also use this technique. Assuming the finite range R of the bounded potential function, we approximate the kernel function as

$$F(x, y) \approx \sum_{k,j=0}^{N} \Delta_k(x) F_{k,j} \Delta_j(y), \qquad (7)$$

where $F_{k,j} \equiv F(kh, jh)$, and the basis functions are

$$\Delta_0(x) = \begin{cases} 0 & \text{for } |x| > h, \\ 1 + x/h & \text{for } -h \leq x \leq 0, \\ 1 - x/h & \text{for } 0 < x \leq h; \end{cases} \qquad (8)$$

$$\Delta_n(x) = \Delta_0(x - hn).$$

where $h$ is some step, and $R = Nh$.

Decreasing the step $h$, one can approach the kernel arbitrarily close at all points. As a result, the kernel is presented in a separable form. We solve eq. (3) by substituting

$$L(x, y) \approx \sum_{k,j=0}^{N} P_k(x) \Delta_j(y). \qquad (9)$$

Substitution of eqs. (7) and (9) into eq. (3), and taking into account the linear independence of the basis functions, gives

$$\sum_{m=0}^{N} \left( \delta_{jm} + \sum_{n=0}^{N} \left[ \int_x^{\infty} \Delta_m(t) \Delta_n(t) dt \right] F_{n,j} \right) P_k(x)$$

$$= -\sum_{k=0}^{N} \Delta_k(x) F_{k,j}. \qquad (10)$$

We need values of $P_k(hp) \equiv P_{pk}$ ($p, k = 0,...,N$). In this case integrals in eq. (10) may be calculated

$$\zeta_{nmp} = \int_{ph}^{\infty} \Delta_m(t)\Delta_n(t)dt =$$

$$= \frac{h}{6}\left(2\delta_{nm}\left(\delta_{np} + 2\eta_{n\geq p+1}\right) + \delta_{n(m-1)}\eta_{n\geq p} + \delta_{n(m+1)}\eta_{m\geq p}\right).$$

Here, along with the Kronecker symbols $\delta_{k,p}$, symbols $\eta_a$ are introduced, which are equal to one if the logical expression $a$ is true, and are equal to zero otherwise. Considering also that $\Delta_k(hp) \equiv \delta_{k,p}$, we finally get the system of equations

$$\sum_{m=0}^{N}\left(\delta_{j,m} + \sum_{n=0}^{N}\zeta_{n,m,p}F_{n,j}\right)P_{pm} = -F_{p,j} \qquad (11)$$

for each $j,p = 0,..,N$.

Solution of (11) gives $P_k(hp) \equiv P_{p,k}$. Potential values at points $r = hp$ ($p = 0,..,N$) are determined from eq. (6) by some finite difference formula.

Next, we consider the case $l = 0$, for which $h_l^+(qx) = e^{iqx}$ and

$$F(x,y) = F(x+y) = \frac{1}{2\pi}\int_{-\infty}^{+\infty} e^{iq(x+y)}Y(q)dq.$$

We approximate the kernel as follows:

$$F(x,y) = F(x+y) \approx \sum_{k=-2N}^{2N} F_{0,k}H_k(x+y), \qquad (12)$$

where $F_{0k} \equiv F(kh)$ as in eq. (7), and the used basis is

$$\left.\begin{array}{l} H_0(x) = \begin{cases} 0 & \text{for } x < 0, \\ 1 & \text{for } 0 \leq x \leq h, \\ 0 & \text{for } x > h; \end{cases} \\ H_n(x) = H_0(x - hn). \end{array}\right\} \qquad (13)$$

The Fourier transform of the basis functions (15) is

$$\tilde{H}_k(q) = \int_{-\infty}^{\infty} H_k(x)e^{-iqx}dx = \frac{i(1-e^{iqh})}{q} \cdot e^{-iqhk}.$$

The function $Y(q)$ is represented in the basis of the functions:

$$Y(q) = \sum_{k=-2N}^{2N} F_{0,k}\tilde{H}_k(q) = \sum_{k=-2N}^{2N} F_{0,k} \cdot \frac{i(1-e^{iqh})}{qh} \cdot e^{-iqhk}.$$

The last relationship may be rewritten as

$$qY(q) = \sum_{k=-2N}^{2N} F_{0,k} \cdot i\left(e^{-iqh} - 1\right) \cdot e^{-iqhk} =$$

$$= i \sum_{k=-2N+1}^{2N} \left(F_{0,k-1} - F_{0,k}\right) e^{-iqhk}$$

$$+ i\left(-F_{0,-2N}\right) e^{iqh 2N} + i\left(F_{0,2N}\right) e^{-iqh(2N+1)}.$$

Thus, the left side of the expression is represented as a Fourier series on the segment $-\pi/h \leq q \leq \pi/h$. Taking into account that $Y(-q) = Y^*(q)$, we get

$$\left. \begin{array}{l} -F_{0,-2N} = \dfrac{h}{\pi} \int_0^{\pi/h} \mathrm{Im}\left(qY(q) e^{-iqh 2N}\right) dq; \\[2mm] F_{0,k-1} - F_{0,k} = \dfrac{h}{\pi} \int_0^{\pi/h} \mathrm{Im}\left(qY(q) e^{iqhk}\right) dq \\[2mm] \text{for } k = -2N+1, \ldots 2N-1; \\[2mm] F_{0,2N} = \dfrac{h}{\pi} \int_0^{\pi/h} \mathrm{Im}\left(qY(q) e^{iqh(2N+1)}\right) dq. \end{array} \right\} \quad (14)$$

The system (14) is solved recursively from $F_{0,2N}$. Thus the range of known scattering data defines the step value $h$ and therefore the accuracy of the inversion.

Calculation results for the potential function $V(r) = -3\exp(-3r/2)$ are presented in Figs. 1, 2. The step $h = 0.04$, $R = 4$. S-matrix was calculated at points shown in Fig. 1 up to $q = 8$. The S-matrix was interpolated by a quadratic spline in range $0 < q < 8$. For $q > 8$ the S-matrix was approximated as asymptotic $S(q) = \exp(-2iA/q)$ for $q > 8$, where $A$ was calculated at $q = 8$.

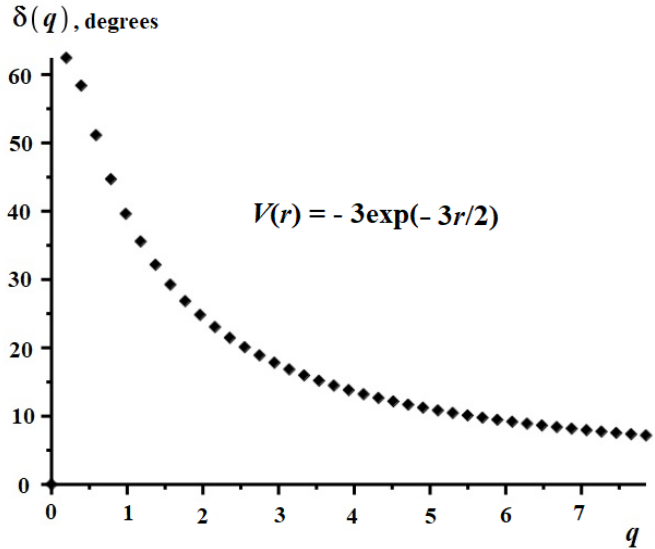

Figure 1. The model scattering data used in calculation. Only shown points were calculated from Eq. (1).

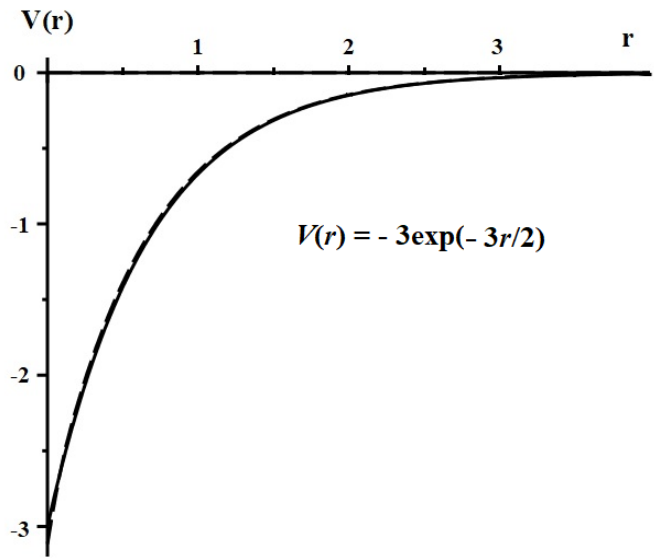

Figure 2. The input potential is shown by solid line and the calculated potential function is indicated by dashed line.

**Conclusions**

Thus, we presented a solution of the quantum scattering inverse problem for zero orbital angular momentum, the algorithm of which is as follows. We set the value of the step $h$, which determines the required accuracy of potential function determination. From the scattering data (2,5), we determine $F_{0,k}$ from (14, 15). Solution of system (11) gives values of $P_k(hp)$ ($p = 0,..,N$). Further, the values of the potential function (6) are determined by some finite difference formula.

Expressions (7-12) give a method for the Marchenko equation's numerical solution for an arbitrary orbital angular momentum $l$, including the case of coupled channels. We will present a generalization of the stated formalism in general for $l > 0$ later after developing a numerical algorithm.